\documentclass[10pt,conference]{IEEEtran}
\IEEEoverridecommandlockouts
\usepackage{cite}
\usepackage{amsmath,amssymb,amsfonts}
\usepackage{algorithmic}
\usepackage{graphicx}
\usepackage{textcomp}
\usepackage{xcolor}
\usepackage{amssymb}
\usepackage{graphicx}
\usepackage{multirow}
\usepackage{graphicx}
\usepackage{array}

\renewcommand{\thefigure}{\arabic{figure}}
\def\BibTeX{{\rm B\kern-.05em{\sc i\kern-.025em b}\kern-.08em
    T\kern-.1667em\lower.7ex\hbox{E}\kern-.125emX}}
\begin{document}

\title{NetMoniAI: An Agentic AI Framework for Network Security \& Monitoring
{\footnotesize \textsuperscript{*} }
\thanks{ }
}
\author{
\IEEEauthorblockN{Pallavi Zambare\IEEEauthorrefmark{1}, Venkata Nikhil Thanikella\IEEEauthorrefmark{2}, Nikhil Padmanabh Kottur\IEEEauthorrefmark{3},Sree Akhil Akula\IEEEauthorrefmark{4} Ying Liu\IEEEauthorrefmark{5} }
\IEEEauthorblockA{\IEEEauthorrefmark{1,2,3,4,5}Department of computer science, Texas Tech University, Lubbock, USA \\
Emails: pzambare@ttu.edu, nikhilvenkata.t@gmail.com, nkotturi@ttu.edu, sreakula@ttu.edu, y.liu@ttu.edu}
 
}

\maketitle

\begin{abstract}
In this paper, we present NetMoniAI, an agentic AI framework for automatic network monitoring and security that integrates decentralized analysis with lightweight centralized coordination. The framework consists of two layers: autonomous micro-agents at each node perform local traffic analysis and anomaly detection. A central controller then aggregates insights across nodes to detect coordinated attacks and maintain system-wide situational awareness. We evaluated NetMoniAI on a local micro-testbed and through NS-3 simulations. Results confirm that the two-tier agentic-AI design scales under resource constraints, reduces redundancy, and improves response time without compromising accuracy. To facilitate broader adoption and reproducibility, the complete framework is available as open source. This enables researchers and practitioners to replicate, validate, and extend it across diverse network environments and threat scenarios. Github link:  https://github.com/pzambare3/NetMoniAI
 \end{abstract}


\section{Introduction}
Network monitoring plays a critical role in detecting threats and maintaining security.\cite{a1} With the expansion of infrastructure across cloud, enterprise, and edge environments, the volume and complexity of network traffic continue to increase. This increase creates more opportunities for attackers and makes real-time visibility more difficult.\cite{a2} Recent cyber incidents highlight the scale and severity of modern threats.  These include state-sponsored attacks during the Russian invasion of Ukraine \cite{b1} and targeted intrusions during the COVID-19 pandemic.\cite{b2}.

To counter these threats, network visibility must be both accurate and scalable. This is achieved through two main approaches to traffic monitoring: packet-based and flow-based. Packet-level analysis offers detailed visibility but often becomes infeasible at scale due to high processing and storage requirements\cite{b3}. Conversely, flow-level monitoring reduces overhead by summarizing traffic into structured records \cite{b4}. This approach is suitable for large-scale environments and is widely adopted in intrusion detection \cite{b5}. However, flow-based methods sacrifice accuracy and timeliness \cite{b6,b7}, which limits their effectiveness in fast-changing threat landscapes. Overcoming this trade-off remains a key challenge in network security.While packet and flow-based monitoring are the foundation of network visibility, most systems still rely on manual decision-making\cite{b4}. Traditional approaches such as static rule engines and centralized log analysis remain common. These approaches are slow to adapt and often generate false positives. Continuing detection quality requires frequent rule updates and operator oversight. This limits scalability and reduces effectiveness against new or distributed threats.\cite{a3} A more autonomous and adaptive approach is needed. 

Several recent approaches have explored introducing automation through machine learning. Quantized autoencoders enable lightweight anomaly detection on IoT devices\cite{b8}. While graph neural networks leverage flow-level topology to capture complex patterns\cite{b9}. Federated and multi-agent systems move inference to the edge to reduce latency \cite{b10,b11}. However, these models still depend on synchronization and lack transparency and do not support autonomous decision-making.\\
\begin{figure}[htbp]
\centerline{\includegraphics[width=7.5cm,height=4.5cm]{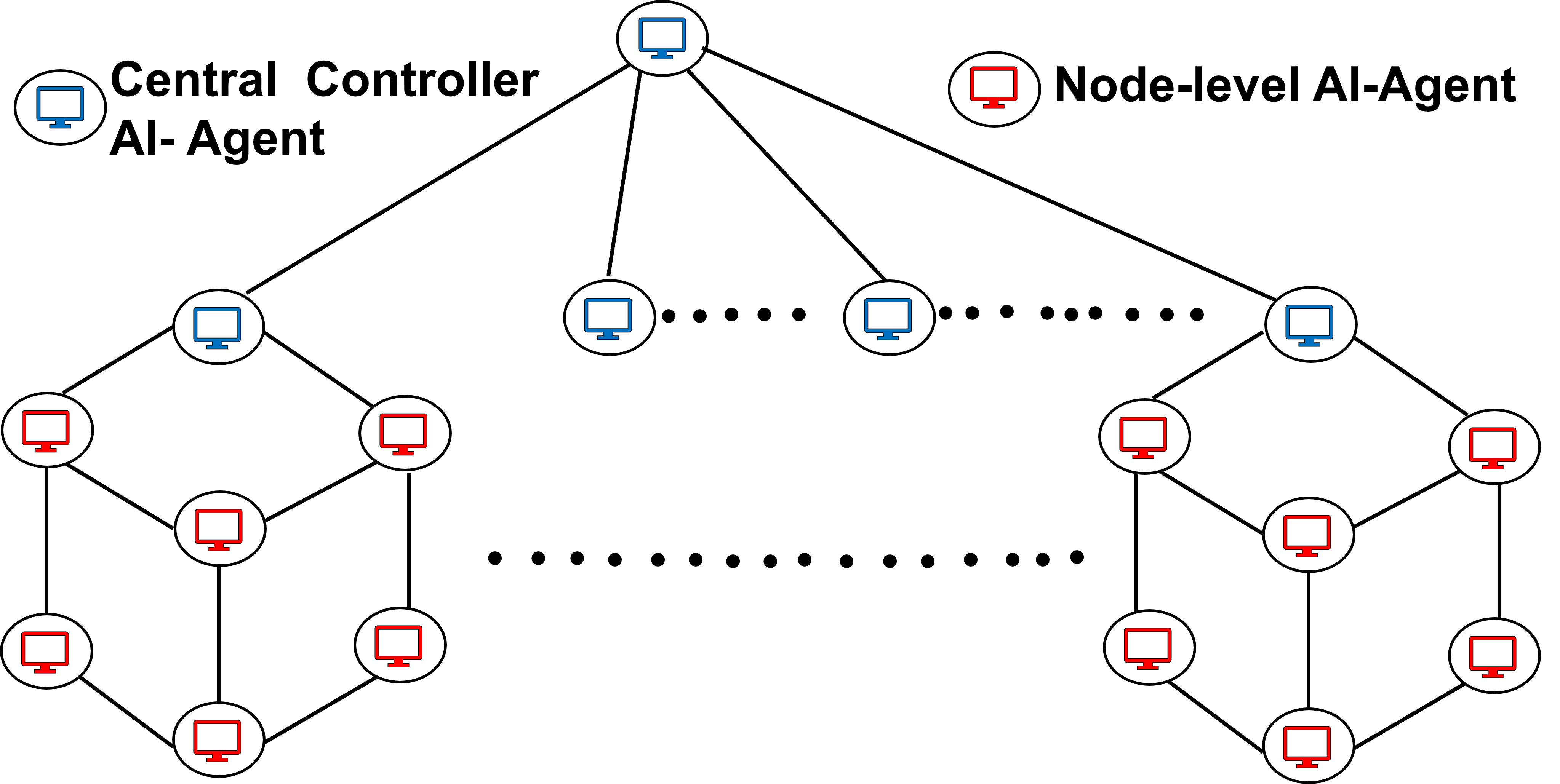}}
\caption{NetMoniAI System Architecture.}
\label{fig1}
\end{figure}
To overcome these limitations, we introduce NetMoniAI, an agentic AI framework for distributed network security. It combines packet-level and flow-level monitoring to support accurate and scalable analysis. Each node runs a lightweight micro-agent. These agents capture traffic, detect anomalies, and reason using LLMs. These agents operate autonomously without requiring static rules or centralized control. A central controller aggregates reports from individual agents to identify cross-node attack patterns and coordinate responses while maintaining local autonomy.  NetMoniAI supports end-to-end functionality, including capturing raw packets, generating human-readable summaries, and correlating alerts across the network. It also performs alert correlation across multiple nodes to support distributed threat detection. The system is evaluated in two distinct scenarios: a real-time degradation test on a Linux-based micro-testbed and a multi-node attack simulation using NS-3.

\section{Proposed Framework}
NetMoniAI is designed as a hybrid Agentic AI framework that combines distributed intelligence at the node level with centralized reasoning at the central level. The system architecture, shown in Figure 1, consists of two primary components. Autonomous Node-Level AI Agents that operate on individual machine and decision making. Centralized Controller AI Agent that coordinates higher-level reasoning and multi-agent integration. This architecture enables both local autonomy and system-wide threat visibility.

\subsection{Node-Level Agent Architecture and Functional Layers}
NetMoniAI agents are lightweight, autonomous microservices deployed at individual network nodes. As illustrated in Figure 2, each agent is organized into four integrated layers that collectively enable end-to-end monitoring, analysis, and decision-making. The Service Layer handles low-level data acquisition and control using modules like the Packet Capture Module, Tuning Controller, and Report Generator. The Agent Layer performs local decision-making using an execution loop composed of the Planner, Tool Executor, and Communication Module. Together, these components collectively manage telemetry processing, semantic reasoning triggers, and internal task routing. The Model Layer connects to external LLMs (e.g., Gemini Pro or GPT-O3) via REST API or executes local BERT models to classify network events. The Application Layer presents results to human operators through a Realtime Dashboard and a Conversational Chatbot interface. Additionally, each agent maintains a lightweight memory buffer to ensure temporally consistent decisions across cycles. This internal memory enables contextual awareness and continuity in threat assessment over time.
\renewcommand{\arraystretch}{1.2}
\begin{table}[!htbp]
\centering
\caption{Component functions across NetMoniAI architectural layers.}
\resizebox{\columnwidth}{!}{%
\begin{tabular}{|c|l|p{5.8cm}|}
\hline
\textbf{Layer} & \textbf{Component} & \textbf{Primary Function} \\
\hline
\multirow{5}{*}{Service}
    & Packet Capture Module   & Collects raw network traffic using tools like \texttt{tshark}. \\
    & Analyzer Engine         & Processes and inspects traffic for anomalies or attack signatures. \\
    & Report Generator        & Converts findings into structured logs or summaries. \\
    & Tuning Controller       & Dynamically adjusts capture duration and sampling intervals. \\
    & Decision Trigger        & Activates downstream actions based on metric thresholds. \\
\hline
\multirow{6}{*}{Agent}
    & Input Handler           & Collects data and alerts from the Service Layer. \\
    & Planner                 & Chooses next steps based on rules or LLM-guided logic. \\
    & Tool Executor           & Executes selected functions or scripts. \\
    & Short-Term Memory       & Temporarily stores state or prior decisions across cycles. \\
    & Execution Loop          & Maintains continuous sensing, decision-making, and acting. \\
    & Communication Bus       & Interfaces with the model layer and external components. \\
\hline
\multirow{2}{*}{Model}
    & LLM Inference Engine    & Performs semantic reasoning on traffic features. \\
    & Function Call Interface & Maps LLM outputs to callable agent actions. \\
\hline
\multirow{2}{*}{Application}
    & Realtime Dashboard      & Displays metrics, anomalies, and system state. \\
    & Conversational Interface & Enables human-LLM interaction through natural-language queries. \\
\hline
\end{tabular}%
}
\label{tab:architecture}
\end{table}

A detailed breakdown of each layer and its associated modules is presented in Table I, which maps architectural layers to their corresponding functional responsibilities within the agent’s workflow. Whereas each agent operates independently, it also contributes structured insights to the broader system. These outputs, generated in both machine-readable and human-interpretable formats, serve as inputs for the central coordination mechanism described in the following section.

\subsection{Central Controller for Multi-Agent Coordination}
The Central Controller in NetMoniAI serves as the centralized reasoning and coordination engine across distributed node agents. While each node-level agent independently detects and interprets local anomalies. These Central Controller synthesizes their outputs to establish system-wide situational awareness and orchestrate coordinated responses. This layered design enables decentralized data collection while maintaining centralized intelligence.\\
\begin{figure}[htbp]
\centerline{\includegraphics[width=8.5cm,height=5.5cm]{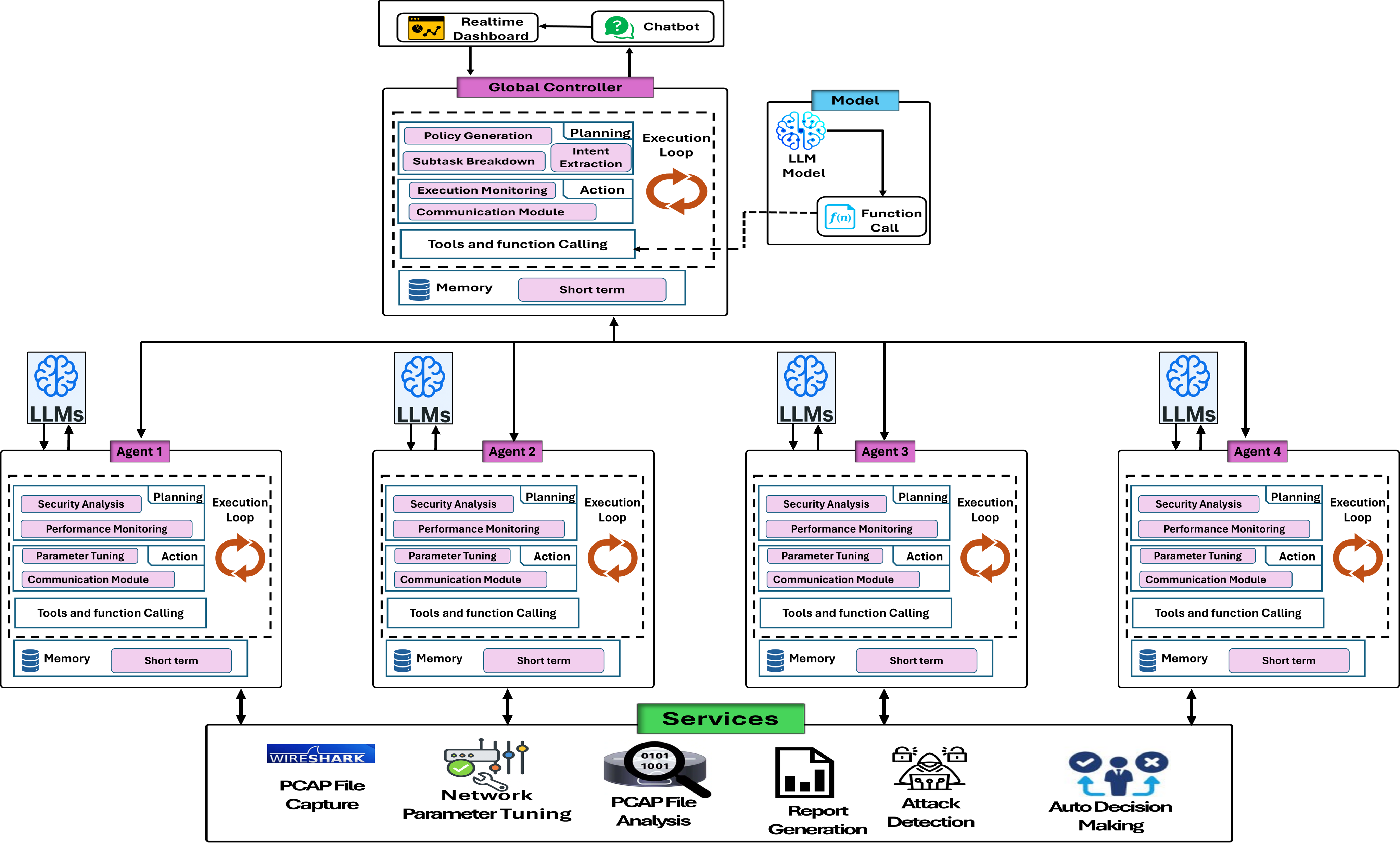}}
\caption{Central Controller Architecture}
\label{fig3}
\end{figure}
As illustrated in Figure 3, the controller is composed of six core modules that operate in a continuous execution loop.  These loop spans planning, action, memory integration, and tool/function invocation. The Planning layer includes Policy Generation, Subtask Breakdown, and Intent Extraction, which collectively convert raw telemetry into high-level goals and contextual understanding. The Action layer consists of Execution Monitoring and a Communication Module, which monitors agent activities and relays synthesized instructions or feedback. Finally, the Memory/Interface Layer maintains recent threat context and facilitates external interaction via a dashboard and chatbot.\\
\begin{table}[!htpb]
\centering
\caption{Core components and functions of the Central Controller  in NetMoniAI.}
\resizebox{\columnwidth}{!}{%
\begin{tabular}{|l|p{9.5cm}|}
\hline
\textbf{Component} & \textbf{Primary Function} \\
\hline
Policy Generator & Defines high-level monitoring goals, thresholds, and system-wide coordination rules. \\
\hline
Subtask Breakdown & Decomposes correlated threats into actionable tasks for system-wide evaluation. \\
\hline
Execution Monitor & Tracks agent health, activity timelines, and identifies delayed or missing updates. \\
\hline
Intent Extractor & Parses agent reports and natural-language summaries to infer attack intentions. \\
\hline
Communication Module & Manages asynchronous message exchange with node agents and dispatches alerts. \\
\hline
Memory Store & Maintains short-term event history for temporal correlation and trend reasoning. \\
\hline
\end{tabular}%
}
\label{tab:central_controller_components}
\end{table}
The controller incorporates a short-term memory buffer to maintain state consistency over time, ensuring that long running or distributed threats are interpreted cohesively. This temporal awareness enhances the controller’s ability to detect evolving patterns. Additionally, the Tool Invocation Engine interfaces with both internal components and external APIs (e.g., LLM-based reasoning or script-based mitigations), enabling adaptive and proactive responses.

Crucially, the Controller is designed in alignment with agentic AI principles. It does not override agent autonomy but instead acts as an advisory and correlation layer. It issues alerts, recommends observation adjustments, or suggests mitigations, leaving final execution to the node-level agents. This design enhances modularity and supports robust coordination even in environments with intermittent connectivity or partial agent failure.
A summary of the controller’s internal modules and their responsibilities is presented in Table II, which complements the architectural view shown in Figure 3.

\section{System Implementations}
NetMoniAI is built using modular microservices that operate independently. These services are developed using modern Python frameworks and follow an asynchronous execution model. This chapter details how each component is technically realized. It explains the services, libraries, communication flows, and semantic reasoning stack that bring the architecture to life. Together, these elements enable event-driven monitoring, context-aware memory, and LLM-powered decision-making.
\subsection{Node-Level Agent Pipeline}
 Each node-level agent is implemented as a FastAPI-based microservice. It mirrors the four-layer architecture described in Chapter 2: Service, Agent, Model, and Application. Internally, each layer runs as an asynchronous coroutine and communicates through asyncio queue structures. This design supports parallel execution of tasks like packet capture, anomaly detection, and semantic reasoning. It ensures fast, lightweight operation even on edge devices.
 
The agent starts by continuously monitoring system metrics using the Performance Monitoring Module. When a threshold is breached, it triggers packet capture through tools like tshark. The raw data is passed to the Feature Extraction Module, which converts packets into a structured CSV format using scapy. This output is sent to the Model Layer, where either a local BERT model or a remote LLM (e.g., Gemini Pro) performs semantic inference to detect and classify threats.

After inference, the results are serialized using Pydantic into structured JSON. These include both threat classifications and human-readable summaries. The Report Generator then sends this information to the central Controller through a REST endpoint. This approach allows downstream systems to process the data efficiently, while also making it understandable to security analysts. Additionally, NetMoniAI agents are not limited to packet-level monitoring. Each agent uses tools like tshark and scapy to inspect raw packets for local anomalies such as delay or jitter. These observations are then aggregated and correlated by the central Controller. This enables the system to interpret flow-level behaviors across the network. The hybrid strategy allows detection of both isolated issues (e.g., SYN floods) and coordinated attacks (e.g., distributed reconnaissance).

\subsection{Central Controller Pipeline}
The Central Controller is implemented as a high-throughput FastAPI service. It receives structured reports and summaries from all node agents using asynchronous REST endpoints. These reports are grouped by source, timestamp, and threat type. Internally, the controller performs correlation, intent extraction, and system-wide threat analysis. It operates independently of local agents, but coordinates responses based on multi-node insights.
The controller includes dedicated modules for aggregation, planning, memory, and communication. The Aggregation Module clusters incoming reports using pandas for in-memory processing. The Intent Extractor compares semantic summaries across agents to infer shared attacker behavior. A short-term memory tracks recent events, allowing the controller to reason over time and detect distributed or stealthy threats. These modules form a continuous loop that supports high-level coordination without interrupting local agent autonomy.
Once a threat is identified, the controller compiles correlated insights into structured reports and natural-language summaries. These outputs are exposed through a WebSocket-based dashboard and a chatbot interface. This design allows real-time system visibility and interactive querying. Importantly, the controller suggests actions but does not issue direct commands. This approach preserves the autonomy of node-level agents while ensuring coordinated, system-wide defense.A summary of the core libraries, tools, and their corresponding versions utilized in the implementation of NetMoniAI is presented in Table III.

\begin{table}[!htpb]
\centering
\caption{Libraries, tools, and versions used in the NetMoniAI.}
\resizebox{\columnwidth}{!}{%
\begin{tabular}{|l|l|p{4cm}|}
\hline
\textbf{Component} & \textbf{Library / Tool} & \textbf{Version / Notes} \\
\hline
Agent framework & Pydantic-AI & 0.110 \\
\hline
Controller framework backend & FastAPI & 2.3 \\
\hline
System telemetry & psutil & Lightweight sampling \\
\hline
Packet capture & tshark / tcpdump & 4.2.x \\
\hline
Feature extraction & scapy & 2.5.0 \\
\hline
Local reasoning model & BERT (Hugging Face) & base-uncased \\
\hline
LLM reasoning & Gemini Pro / GPT-o3 & API-based inference \\
\hline
Frontend dashboard & ReactJS + WebSocket & Node.js 20.x \\
\hline
Message typing & Pydantic & Enforces schema validity \\
\hline
\end{tabular}%
}
\label{tab:component_versions}
\end{table}
\subsection{Deployment Environments}
NetMoniAI was deployed and validated under two distinct settings to assess its robustness, responsiveness, and adaptability in realistic conditions.\\
\begin{figure}[htbp]
\centerline{\includegraphics[width=8.5cm,height=5.5cm]{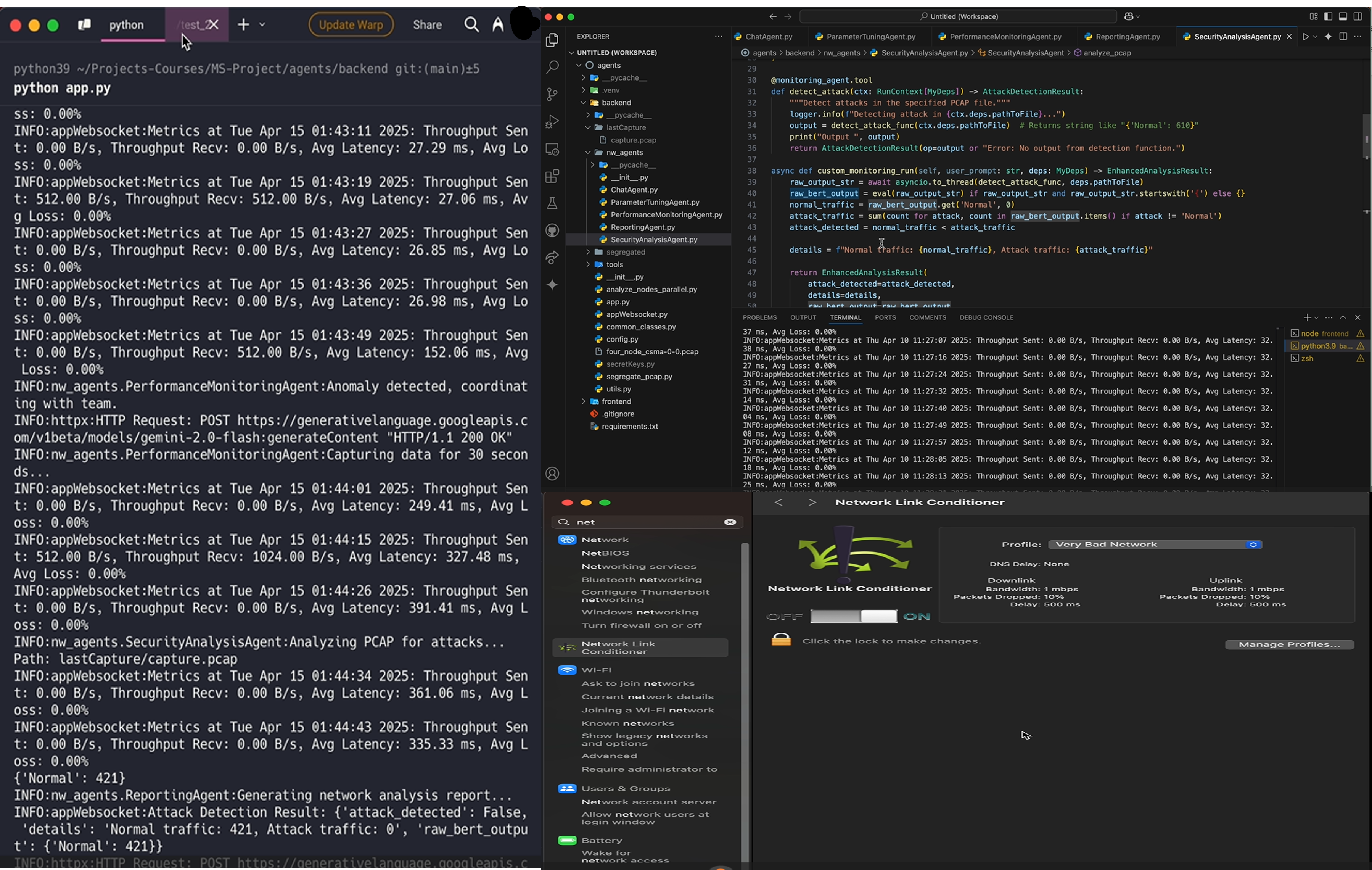}}
\label{figtestbed}
\caption{Local micro-testbed configuration for performance validation.
The setup simulates varied link conditions across Baseline, Degraded, and Recovery phases using controlled latency and bandwidth constraints.
}
\end{figure}
\subsubsection{Local Micro-Testbed}
A controlled testbed was established on a single Ubuntu 22.04 machine to emulate varied network conditions. Network behavior was modulated using the Linux tc utility alongside macOS’s Network Link Conditioner, as shown in Figure 4. This setup enables the emulation of three distinct operational phases: Baseline, Degraded, and Recovery. The “Very Bad Network” profile imposed a consistent 600ms delay and limited both uplink and downlink bandwidth to 1Mbps to simulate degraded link performance. Within this environment, the NetMoniAI agent continuously monitored key metrics such as latency, jitter, and packet loss. Upon crossing predefined thresholds, it initiated packet capture using tshark. It then invoked the GPT-O3 LLM for semantic threat inference and successfully identified abnormal traffic patterns indicative of potential attacks. Structured outputs were generated in both JSON and human-readable summaries. Across repeated tests, the system consistently achieved detection latencies under 5 seconds. This demonstrates its responsiveness and reliability in edge-like constraints.
\subsubsection{NS-3 Simulation Environment}
To evaluate NetMoniAI at scale, a virtual network topology with up to 50 nodes was simulated using the NS-3 framework. This environment enabled deterministic control over network behavior, including traffic patterns, delays, and link characteristics. As a result, it was ideal for reproducible and fine-grained testing. Multiple attack scenarios were emulated using custom traffic scripts, including UDP-based denial-of-service (DoS), targeted single-node flooding, and port scanning. Each simulated node independently captured traffic. Embedded NetMoniAI agents on these nodes conducted semantic analysis of the captured data. The resulting structured reports were then transmitted to the central Controller for correlation and centralized threat detection.This setup tested the framework's ability to detect both localized and distributed attacks. For example, during a coordinated reconnaissance attack, local agents flagged abnormal probing patterns. The central Controller correlated these events across nodes to detect a broader threat. This confirmed that NetMoniAI’s dual-layer architecture can detect both immediate and stealthy threats across a large-scale, dynamic networks.
\section{Results and Evaluation}
This chapter presents the experimental outcomes of deploying NetMoniAI in both controlled and large-scale environments. The evaluation focuses on key performance indicators such as detection latency, responsiveness under degraded conditions, and the system's ability to identify both localized and distributed attacks. Through empirical testing, the effectiveness of NetMoniAI’s agentic architecture was validated. The goal was to assess its ability to deliver accurate, timely, and interpretable threat intelligence across diverse network scenarios. To validate NetMoniAI’s performance, we conducted a series of controlled experiments that evaluated its responsiveness, accuracy, and scalability. The system was tested in both a real-world environment and a simulated multi-node setup. This approach enables assessment of localized detection and central coordination capabilities. Due to space constraints, only selected results are presented here; the complete experimental dataset and visualizations are available in the project's GitHub repository.
\subsection{Local Testbed Evaluation}
This evaluation was conducted under the Degraded phase of the local micro-testbed described in Section 3.3. In this case, the network link was artificially constrained to 1Mbps and experienced a consistent 600ms delay. These controlled impairments provided a realistic scenario for assessing NetMoniAI's responsiveness under stress.\\
\begin{figure}[htbp]
\centerline{\includegraphics[width=8.5cm,height=4.5cm]{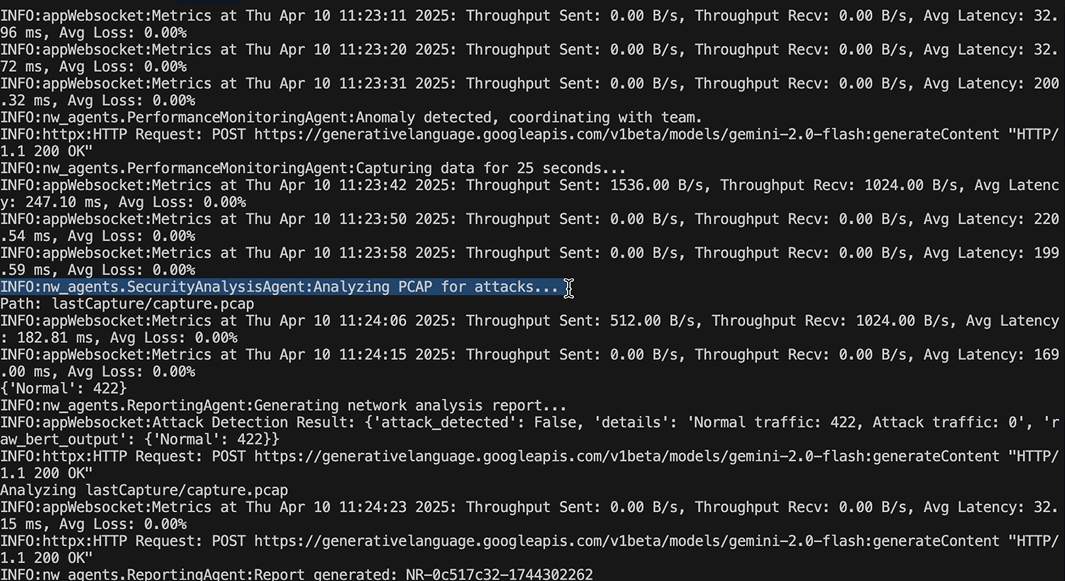}}
\caption{Agent log showing anomaly detection, PCAP analysis, and benign traffic classification using GPT-O3}
\label{figr2}
\end{figure} 
During a representative run, the NetMoniAI agent detected a latency anomaly exceeding 200ms at timestamp 11:23:31, as shown in Figure 5. The Performance Monitoring Module triggered the detection pipeline, initiating a 25-second packet capture using tshark. The captured traffic was parsed using Scapy and analyzed by the Security Analysis Module, which submitted features to the GPT-O3 language model via a structured prompt. The model classified the session as benign, identifying 422 normal packets and no indicators of malicious behavior.

Subsequently, the Reporting Module generated structured JSON and natural-language summaries. These were made available through the REST API and visualized in real time via the dashboard’s WebSocket interface. This confirmed that NetMoniAI was capable of performing complete anomaly detection, LLM-powered reasoning, and interpretable reports.All of this was achieved within 5 seconds of detecting the anomaly, relying solely on the node-level agent pipeline.

 \begin{figure}[htbp]
\centerline{\includegraphics[width=8.5cm,height=4.5cm]{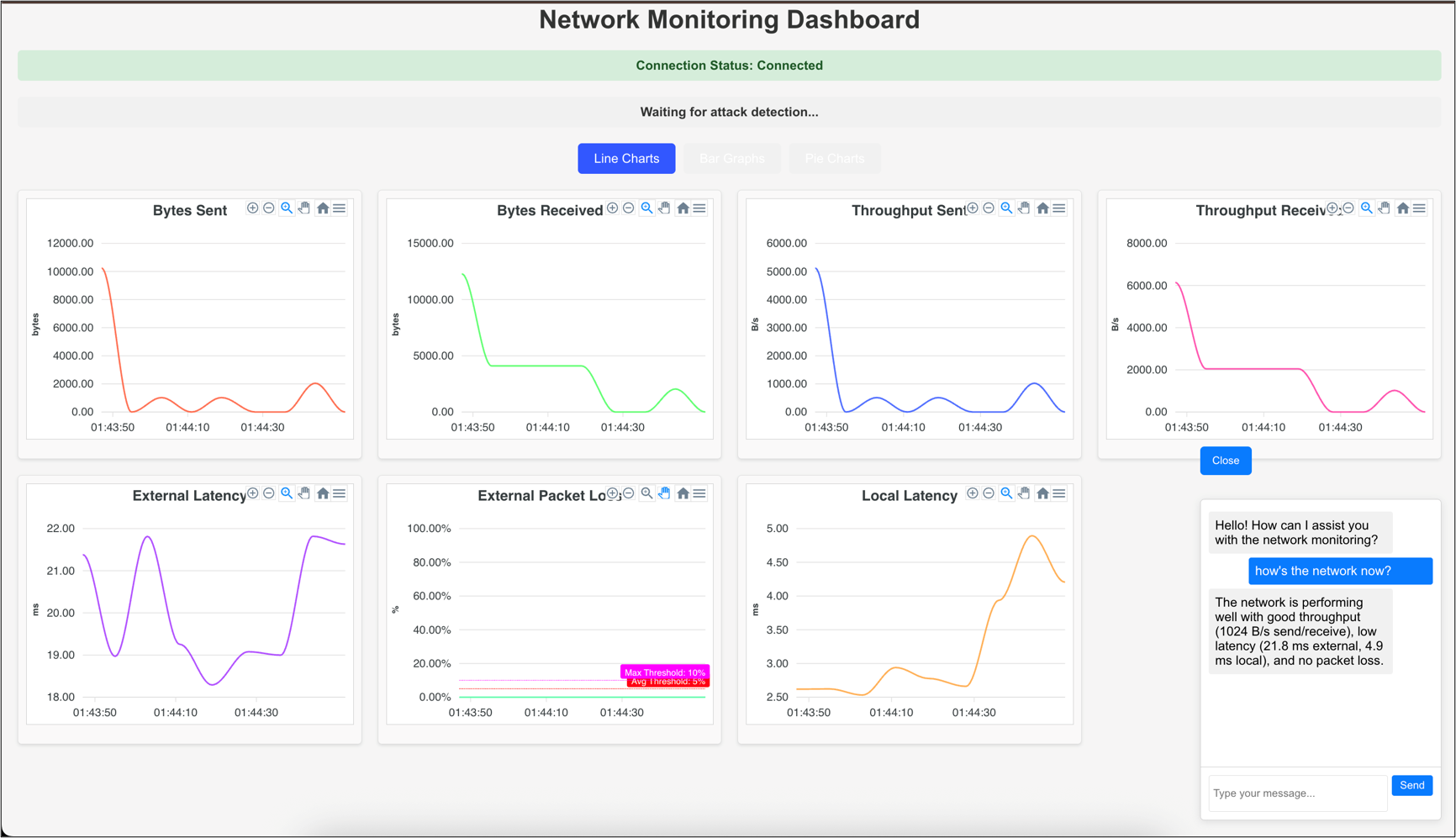}}
\caption{Dashboard view under baseline conditions.}
\label{figlcd1}
\end{figure}
Figures 6 and 7 illustrate the dashboard’s performance during this test. In the baseline phase (Figures 6), metrics such as throughput, latency, and packet loss remained within normal bounds, while the chatbot confirmed healthy status. During the degraded phase (Figure 7), the dashboard reflected sharp increases in both local and external latency. These anomalies were accompanied by threshold breach alerts and updated LLM-generated summaries. This experiment validates that NetMoniAI delivers real-time detection and contextual interpretation, even under degraded network conditions. The dashboard’s interpretability and responsiveness further reinforce its suitability for edge deployments and operational monitoring.
 \begin{figure}[htbp]
\centerline{\includegraphics[width=8.5cm,height=4.5cm]{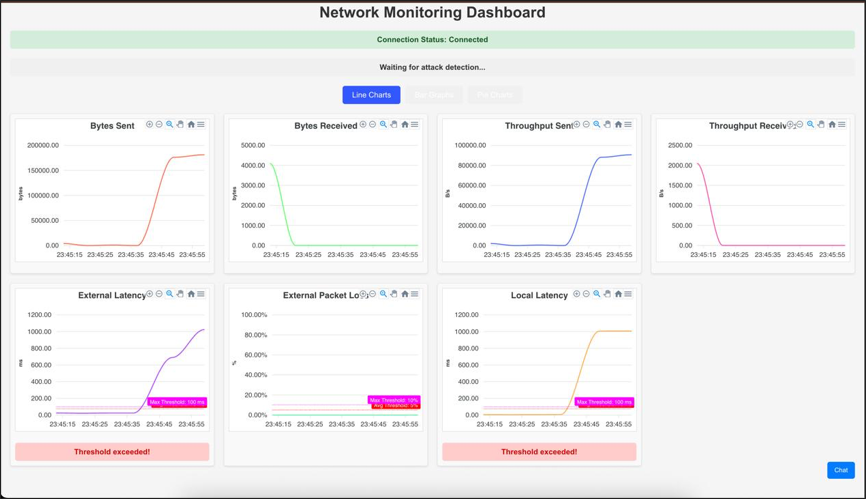}}
\caption{Dashboard view under degraded conditions.}
\label{figlcd2}
\end{figure}
\subsection{NS-3 Simulation-Based Evaluation}
To assess the scalability and distributed intelligence of NetMoniAI, the system was deployed in a virtual environment using the NS-3 network simulator. A reproducible topology was created with eight nodes positioned randomly within a 100×100-meter area. Each node was configured with static IP addresses ranging from 192.168.1.1 to 192.168.1.8. The simulation lasted ten seconds and incorporated both benign background activity and a coordinated TCP flood attack. This setup was designed to emulate a realistic distributed denial-of-service (DDoS) scenario. Background communication was simulated using UDP echo traffic between two node pairs. Simultaneously, Node 1 launched a TCP flooding attack targeting Nodes 4 and 6, both of which hosted a vulnerable TCP packet sink service. The attack was scheduled to begin at 1 second and continued until 9 seconds, with the attacker generating 10 parallel streams of 512-byte packets at a rate of 500kbps per stream. Packet capture tracing was enabled in promiscuous mode on all nodes. This allows each embedded NetMoniAI agent to monitor traffic independently and perform real-time anomaly analysis as visualized in Figure 8.
\begin{figure}[htbp]
\centerline{\includegraphics[width=8.5cm,height=4.5cm]{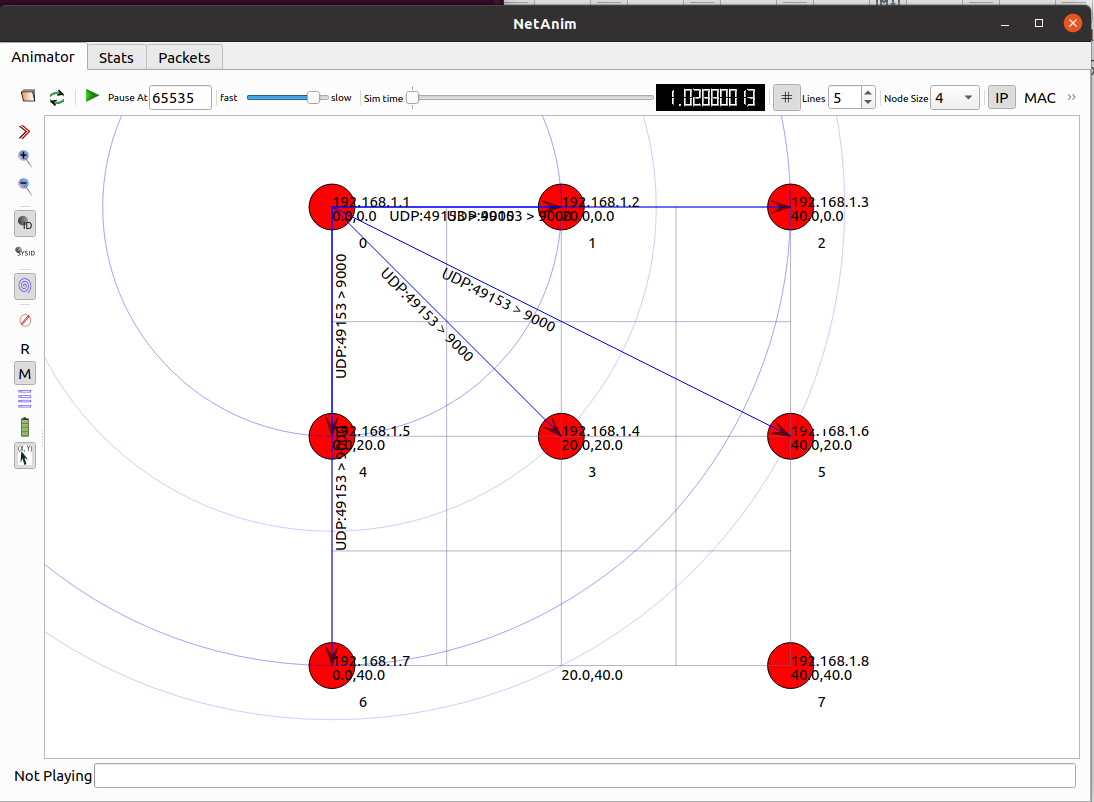}}
\caption{NS-3 Attack Scenario Visualization through NetAnim, illustrating simulated DDoS traffic flows across distributed nodes.}
\label{fignetanim}
\end{figure}
Each NetMoniAI agent autonomously conducted local anomaly detection using real-time traffic features. These structured inferences were transmitted to the central Controller. The Controller aggregated the reports from all nodes to perform cross-node analysis. It examined key factors such as flow direction, transmission timing, and traffic frequency. Based on this analysis, the Controller accurately identified a coordinated multi-source intrusion. It flagged Node 1 as the attacker, and Nodes 4 and 6 as the primary victims. This experiment validated NetMoniAI’s ability to detect DDoS attacks by combining decentralized monitoring with centralized semantic reasoning.
\begin{figure}[htbp]
\centerline{\includegraphics[width=8.5cm,height=4.5cm]{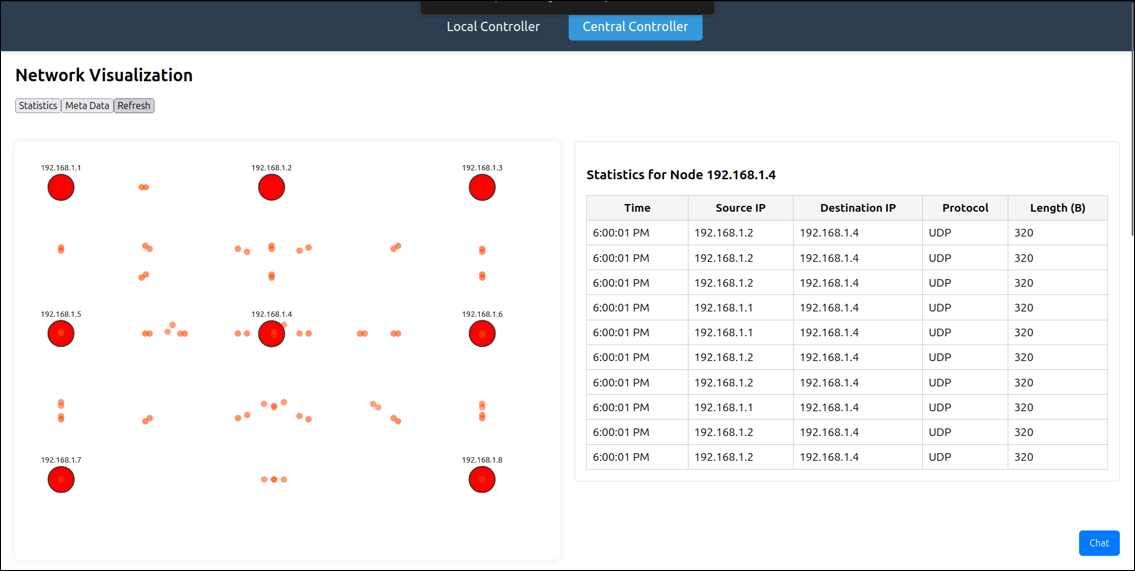}}
\caption{Network Visualization of Real-Time Node Interactions}
\label{figgcd1}
\end{figure}

To support real-time interpretability, the central Controller’s dashboard visualizes node-level behaviors and inferred roles during attack scenarios. As illustrated in Figures 9-12, the dashboard distinctly highlights attackers through outbound traffic patterns. The victim nodes are identified through asymmetric inbound loads. These visual cues provide an intuitive overview of network anomalies. when combined with LLM-generated policy recommendations, enable clear identification of coordinated threats and affected nodes without manual inspection.
\begin{figure}[htbp]
\centerline{\includegraphics[width=8.5cm,height=4.5cm]{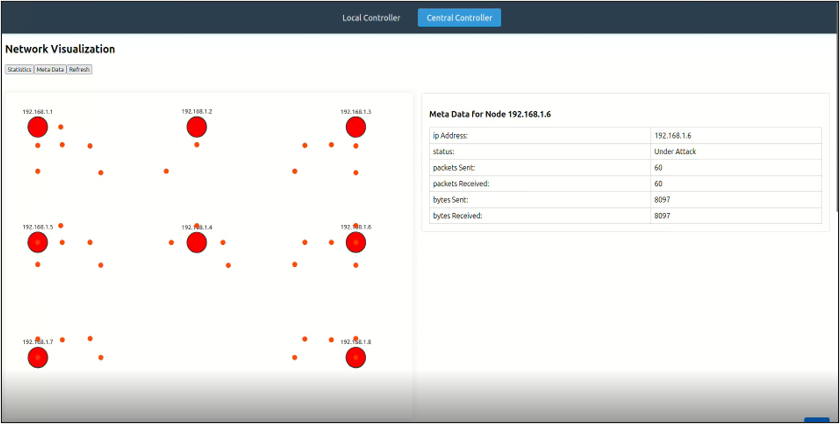}}
\caption{Meta View of Attacked Node During DDoS Scenario}
\label{figgcd2}
\end{figure}

\begin{figure}[htbp]
\centerline{\includegraphics[width=8.5cm,height=4.5cm]{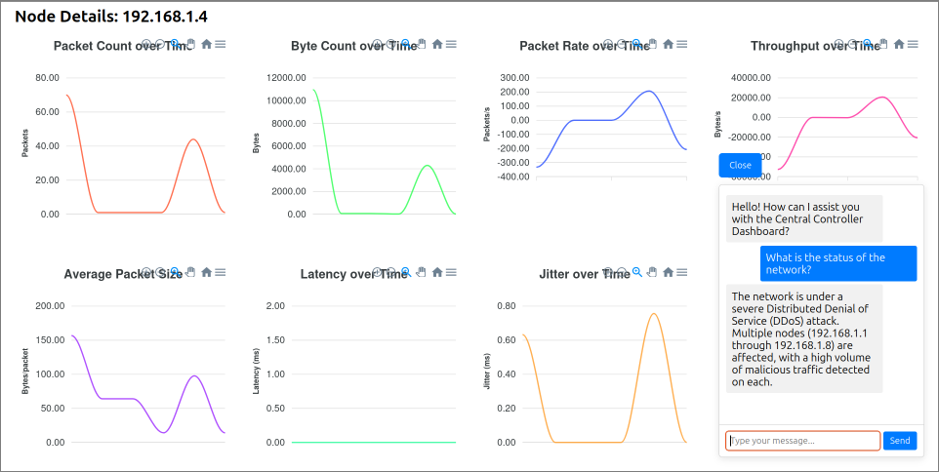}}
\caption{Node 192.168.1.4 Traffic Analysis and Chatbot Insight}
\label{figgcd3}
\end{figure}

\begin{figure}[htbp]
\centerline{\includegraphics[width=8.5cm,height=4.5cm]{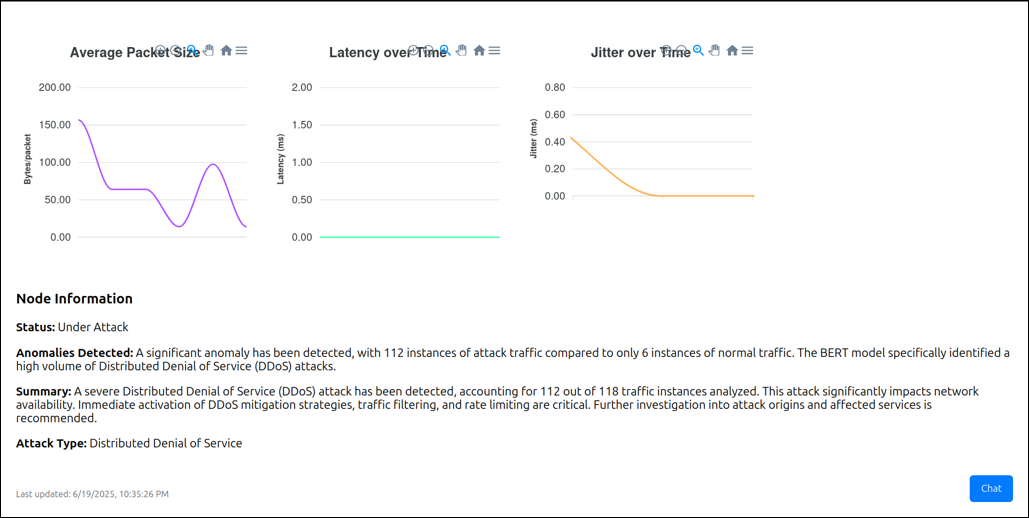}}
\caption{DDoS Attack Summary for Node 192.168.1.4}
\label{figgcd4}
\end{figure}
These insights were reflected in real time on the dashboard, accompanied by LLM-generated policy recommendations for mitigation. Overall, the NS-3 simulation confirmed the effectiveness of NetMoniAI’s hybrid agent-controller architecture. The system demonstrated scalable, distributed threat detection, dynamic role classification, and responsive semantic analysis. Particularly, it achieved these capabilities without introducing processing bottlenecks or significant latency overhead.

\subsection{Conclusion}
This paper presented NetMoniAI, a hybrid agentic AI framework for real-time, distributed network monitoring and threat detection. By combining decentralized sensing at node level with centralized semantic analysis using GPT-O3, the system detects both localized and coordinated attacks with low latency and high accuracy. Evaluated across a local micro-testbed and NS-3 simulations, NetMoniAI demonstrated timely anomaly detection, accurate DDoS classification, and clear operator feedback through structured reports and an interactive dashboard. Its scalable, asynchronous architecture supports interpretable, layered responses without sacrificing performance. Future work will extend the framework with adaptive mitigation, multi-agent coordination, and SDN-based policy enforcement.

\end{document}